\documentclass{mn2e}
\usepackage{epsfig,natbib2,natbibmnfix}

\newcommand{\be}{\begin{equation}}
\newcommand{\ee}{\end{equation}}

\newcommand{\apj}{ApJ}

\newcommand{\mnras}{MNRAS}
\newcommand{\aap}{A\&A}

\newcommand{\apjl}{ApJL}

\newcommand{\nat}{Nature}

\def\ltsima{$\; \buildrel < \over \sim \;$}
\def\simlt{\lower.5ex\hbox{\ltsima}}
\def\gtsima{$\; \buildrel > \over \sim \;$}
\def\simgt{\lower.5ex\hbox{\gtsima}}
\newcommand\sgra{Sgr~A$^*$}

\newcommand\mcapt{\dot{M}_{\rm capt}}

\def\del#1{{}}

\def\msun{{\,{\rm M}_\odot}}
\newcommand\mbh{{\,{\rm M}_{\rm bh}}}

\title{Constraining the number of compact remnants near Sgr~A$^*$}

\author[P.~Deegan \& S.~Nayakshin] {\parbox{18cm}{Patrick Deegan and
Sergei Nayakshin}\vspace{0.3cm}\\ Dept. of Physics \& Astronomy,
University of Leicester, Leicester, LE1 7RH, UK}

\begin{document}

\maketitle

\begin{abstract}
Due to dynamical friction stellar mass black holes and neutron stars are
expected to form high density cusps in the inner parsec of our Galaxy.  These
compact remnants, expected to number around 20000, may be accreting cold dense
gas present there, and give rise to potentially observable X-ray
emission. Here we build a simple but detailed time-dependent model of such
emission. We find that at least several X-ray sources of this nature should be
detectable with {\em Chandra} at any one time. Turning this issue around, we
also ask a question of what current observational constraints might be telling
us about the total number of compact remnants. A cusp with $\sim 40$ thousand
black holes over-predicts the number of discrete sources and the total X-ray
luminosity of the inner parsec, and is hence ruled out. Future observations
of the distribution and orbits of the cold ionised gas in the inner parsec of
\sgra\ will put tighter constraints on the cusp of compact remnants.
\end{abstract}

\renewcommand{\thefootnote}{\fnsymbol{footnote}}
\footnotetext[1]{E-mail: {\tt Sergei.Nayakshin at astro.le.ac.uk}}

\section{Introduction}
\label{intro}

Theoretical calculations predict a cusp of around $\sim$ 20000 stellar mass
black holes in the central parsec of our Galaxy
\citep{Morris93,Miralda00} and a similar number of neutron
stars \citep{Freitag06,Hopman06}. X-ray observations reveal a highly significant
overabundance of transients in the same region \citep{Muno05} as compared with
the region of $\sim$ several tens of parsecs from \sgra, the $M_{\rm smbh}
\sim 4 \times 10^6 \msun$ super-massive black hole \citep{Schoedel02,Ghez03a}
in the Galactic Centre (GC). \cite{NS07} proposed that these compact remnants
may be accreting gas at relatively high rates when they happen to travel
through a dense ionised gas observed to exist in the GC \citep[also
see][]{Morris93}. They calculated a simple time-averaged model for X-ray
emission from such a cusp, and concluded that the total emission of the cusp
could be as high as $\simgt 10^{35}$ erg s$^{-1}$, i.e. very significant
observationally.

On the other hand, a time-independent treatment does not do justice to the
complexity of the problem. Despite the high total number of black holes, due
to a small volume filling fraction of cold gas in the GC, only a few of the
black holes will be moving within the gas clouds and possess a small enough
relative velocity to be visible to {\em Chandra}. Here we extend the model of
\cite{NS07} in two ways. Firstly, we allow time-dependency in the problem by
explicitly following realistic Keplerian orbits of the compact
objects. Secondly, we model the formation and evolution of small scale
accretion discs around the accretors, as such discs will surely form due to an
excess angular momentum in the accreting gas. Non-circular gas orbits are also
considered with a simplified approach.

Our models show a large intrinsic time-dependence of the accretion on the
compact objects and the X-ray emission it produces. Despite that, and despite
internal uncertainties of the model (exact gas orbits, circularisation radius
parameter, radiative efficiency, etc.), we feel that certain rather robust
conclusions can be drawn. In particular, with the $\sim 20000$ compact
remnants expected in the central parsec, at least several X-ray sources with
X-ray luminosity greater than $10^{33}$ erg s$^{-1}$ should be present. Such
sources, which can be called ``fake X-ray binaries'' for obvious reasons, can
potentially contribute to the sources observed by \cite{Muno05} in the central
parsec. Conversely, it appears that a cusp significantly more populous, i.e.,
with 40000 compact remnants, would over-produce the X-ray emission as compared
to the observations, and should thus be ruled out.

\section{Numerical approach}\label{sec:method}

The basis of our setup is explained in \cite{NS07}, and hence we shall only
briefly discuss the main points, sparing detail for the improvements
introduced here.

We assume that a stellar mass black hole travelling through a gas cloud or a
disc with density $\rho$ is capturing gas in a {\em small scale disc about
it} (see below) at the \cite{Bondi44} accretion rate:
\begin{equation}
\mcapt = 4 \pi \rho \frac{(G \mbh)^2}{(\Delta v^2 + c_{\rm s}^2)^{3/2}}\;,
\end{equation}
where $c_{\rm s}$ and $\Delta v$ are the gas sound speed and the relative
velocity between the black hole and the gas, respectively.  The above picture
is complicated by the presence of the SMBH. The area of influence of the
stellar mass black holes is limited by the Hill's radius, $ r_{\rm H} = R
(M_{\rm bh}/3M_{\rm smbh})^{1/3}$, where R is the distance between the stellar
mass black hole and the SMBH. This imposes a limit on the capture rate, given
by the Hill accretion rate, $\dot{M_{\rm H}} = 4\pi r^2_{\rm H}\rho c_{\rm
s}$. Hence,
\begin{equation} \dot{M}_{\rm capt} = min[\dot{M}_{\rm capt},\dot{M}_{\rm H}]\end{equation}
where $\dot{M}_{\rm capt}$ on the right hand side of the equation is defined in
eq.1.

\subsection{Time-dependent disc accretion}\label{sec:smalldisc}

The captured gas may have a net angular momentum resulting in the formation of
a disk around the stellar mass black hole. The disc size is of order of the
circularisation radius for the gas flow, $r_{\rm c}$, which is unknown a
priori.  The maximum value of $r_{\rm c}$ is the capture radius, $r_{\rm
capt}$, which is
\begin{equation} 
r_{\rm capt} = min\left[r_{\rm H},\frac{GM_{\rm bh}}{\Delta v^2 + c^2_{\rm
      s}}\right]\;.
\label{rcapt}
\end{equation}
We thus parametrise the circularisation radius as
\begin{equation}
r_{\rm c} = \zeta r_{\rm capt}\;,
\end{equation}
where $\zeta$ is a parameter less than unity. The viscous time scale in such a
disk around the stellar mass black hole is
\begin{equation} 
t_{\rm visc} = \frac{1}{\alpha \Omega_{\rm d}}\left(\frac{r_{\rm
capt}}{h}\right)^2\;,
\end{equation}
where $\alpha$ is the viscosity parameter, $\Omega_{\rm d} = \sqrt{GM_{\rm
bh}/r_c^3}$ is the angular velocity of the disk and $h$ is the height scale of
the disk. Numerically,
 \begin{equation}
t_{\rm visc} = 1.5 \times 10^3 \; \hbox{year} \; \alpha_{0.01}^{-1} \mu_{\rm d}
    r_{\rm c, 12}^{1/2} \; T_{\rm d, 3}^{-1}\;,
\label{tvisc}
\end{equation}

where $T_{\rm d, 3}$ is the disk temperature in units of $10^3$ K, the
viscosity 
parameter is $\alpha = 0.01 \alpha_{0.01}$, $\mu _{\rm d}$ is the mean
molecular 
mass in units of Hydrogen mass and $r_{\rm c, 12}$ is the
circularisation radius in $10^{12}$ cm. This is to be compared with period
(the orbital time) about the SMBH:
\begin{equation}
P = \frac{2\pi}{\Omega_{\rm K}} = 2900 \; \hbox{year}\; R_{0.1}^{3/2} M_6^{-1/2}\;.
\end{equation}
Here, $\Omega_{\rm K}$ is the Keplerian angular frequency for the black hole
orbiting the super-massive one: $\Omega_{\rm K} = \sqrt{GM_{\rm
smbh}/R^3}$. The respective Keplerian velocity is $v_{\rm K} = R \Omega_{\rm
K}$.

Thus, the gas captured in the small scale disk accretes on the black hole
after a delay of a fraction of to a few (black hole around the SMBH) orbital
times.  The evolution of the disk mass is given by the rate at which the mass
is added, $\dot{M}_{\rm capt}$, minus the mass accreted onto the black hole,
$\dot{M}_{\rm acc}$:
\begin{equation}
\frac{d M_{\rm d}}{d t} = \dot{M}_{\rm capt} - \dot{M}_{\rm acc}.
\end{equation}
The black hole accretion rate is calculated as
\begin{equation} 
\dot{M}_{\rm acc} = \frac{M_d}{t_{\rm visc}}
  e^{-t_{\rm visc}/12t}\;
\end{equation}
(see \S 5.2 in \cite{Frank02}). Finally, the luminosity of the accretion flow
is modelled in the same way as in \cite{NS07}. Namely, we write $L_{\rm X} =
\epsilon \dot{M}_{\rm acc} c^2$, where $\epsilon$ is given by
\begin{equation}
\epsilon=\ 0.01 \frac{\dot{M}_{\rm acc}}{\dot{M}_0+\dot{M}_{\rm acc}}\;,
\label{eps}
\end{equation} 
where $\dot M_0=0.01$ is the critical accretion rate where the switch from the
radiatively efficient to radiatively inefficient regime occurs \citep{Esin97}.
Here we also assumed that X-ray emission visible in the {\em Chandra} band
constitutes 10\% of the bolometric efficiency, which would be a lower limit
for typical spectra of X-ray binaries in their hard state.

\subsection{Orbital evolution of accretors}\label{sec:orbits}

We model the velocity and space distribution of stellar mass black holes as a
cusp that follows the \cite{Bahcall76} distribution for heavier species in a
mass-segregated cusp. This distribution results in the black hole number
density and velocity distribution obeying power laws of the form $R^{-7/4}$
and $f_0$ respectively, where,
\begin{equation}f_0 \propto R^{-1/4}\left(1-\frac{v^2}{v_{\rm
      esc}^2}\right)^{-1/4}\;.
\end{equation}
Both space and velocity distributions are isotropic. The most recent
Monte-Carlo simulations \cite{Freitag06} broadly support these classical
results. A recent Fokker-Planck study \citep{Hopman06} predicts a somewhat
steeper power-law density dependence for the black hole cusp, $\rho_{\rm BH}
\propto R^{-2}$, but we leave this level of detail for future
investigations. 

Our black hole cusp is sharply cut at $R = 0.7$ pc \citep[see
][]{Miralda00}. Clearly, the artificial cut of the black hole distribution at
the outer cusp radius is a crude approximation to the more complicated broken
power-law structure of the cusp found by \cite{Freitag06}. It is also not
entirely self-consistent as black holes do follow their orbits hence changing
their radial position. To determine the significance of this we examined the
structure of our model cusp in $10^4$ years after it was set up. The density
profile did in fact change. Some of black holes on eccentric orbits with large
semi-major axes were found at radii much larger than the outer cusp radius
assumed, as expected. However, the maximum change in the black hole density
profile was no more than 30\%. Furthermore, the black holes on very eccentric
orbits will be those that accrete gas at a low rate unless gas in the inner
parsec moves on similar eccentric orbits. Therefore emission from these black
holes might be neglected in any event.

Generating a series of orbits consistent with this space-velocity
distribution, we randomly set the initial phases of the black holes along
their orbits, and then we follow their spatial motion. We also track the
instantaneous gas capture rate for each black hole. When one of these orbits
intersects the disc of the Minispiral, the black hole in question starts
capturing gas and builds up a disc around it as described in \S
\ref{sec:smalldisc}.

\subsection{The model for the Minispiral}\label{sec:mini}

\cite{Paumard04} suggests that the Minispiral is a dynamical feature in a
state of almost a free fall onto \sgra. We feel that the Minispiral's orbit is
not likely to be so radial, as one would then expect \sgra\ itself to be
accreting from the Minispiral. This would result in accretion rates far above
that from the stellar winds \citep{Cuadra06}, and would contradict the X-ray
observations \citep[e.g.,][]{Baganoff03a}. More realistically the gas in the
Minispiral follows an eccentric orbit which does not enter the inner arcsecond
($\sim 0.03$ parsec) of the GC.

In our simple model, the Minispiral is modelled as half of a disc in a local
Keplerian circular rotation around \sgra\ with the total gas mass of $M_{\rm
disc} = 50 M_{\odot}$, in accord with estimates in \cite{Paumard04}. It
extends from a radius of 0.1 pc from the SMBH to a radius of 0.5 pc. The disc
heightscale, $H$, is assumed to have a fixed ratio to the radius, $R$:
$H/R=0.1$. The gas density is given by $\rho(R) = M_{\rm disc}/(\pi R^2 H)$.

The dynamical age of the Minispiral is a few thousand years. Therefore, we ran
our calculations for 3000 years with these assumptions, and then we
``remove'' the Minispiral instantaneously. This is done as a rough model of
time evolution of the system in the case the gas apocenter is larger than 0.5
parsec, so that the Minispiral would leave the inner 0.5 parsec after a
dynamical time.  In \S \ref{sec:sens} we vary some of the above assumptions
about the structure of the Minispiral to estimate sensitivity of our results
to these assumptions.  In a future work a more complicated, but unavoidably
model-dependent dynamic of the Minispiral should be included.

\section{Results}\label{sec:results}

\subsection{Emission from individual black holes}\label{sec:individual}

\begin{table}
\caption{Orbital parameters of individual black hole orbits (see \S
  \ref{sec:individual}). The inclination of the orbit is with respect to the
  midplane of the Minispiral. The last column shows the time-averaged X-ray
  luminosity of the source.}
\begin{center}
\begin{tabular}{|c|c|c|c|c|}\hline
Black  & Inclination & Eccentricity & Semi-major & $<L_{\rm x}>$\\ 
hole & & &axis &\\
 & ($^o$) & & (pc)&erg s$^{-1}$\\ \hline
                                  \hline
T1 & 6 & 0.1 & 0.1& $1.78\times10^{36}$\\ \hline
T2 & 11 & 0.2 & 0.1& $2.97\times10^{34}$\\ \hline
T3 & 1 & 0.6 & 0.3& $1.34\times10^{34}$\\ \hline
T4 & 29 & 0.5 & 0.08& $1.28\times10^{32}$\\ \hline
\end{tabular}
\end{center}
\end{table}

\begin{figure}
\begin{center}
\epsfig{figure=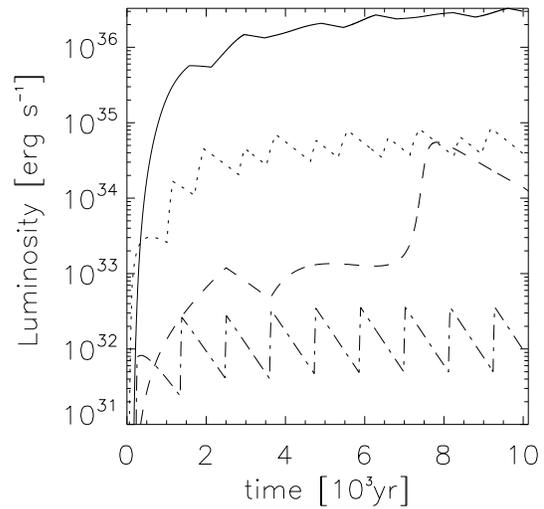}
\caption{X-ray light curves of the four individual black holes described in \S
 \ref{sec:individual}; the solid, dotted, dashed and dotted-dashed lines
 correspond to T1, T2, T3, and T4 respectively.}
\label{fig:1}
\end{center}
\end{figure}

To motivate the study of X-ray emission from black hole cusps in this paper,
we first examine the emission from individual black holes. For simplicity of
discussion in this section only, the ``half-disk'' described in \S
\ref{sec:mini} is replaced with a full disk, with other parameters unchanged,
the only exception being the mass of the Minispiral, which was doubled.  The
black holes follow Keplerian orbits (\S \ref{sec:orbits}) that are
characterised by the values of the semi-major axis and the eccentricity (see
Table 1). The inclination of the orbit to the midplane of the disc, $i$, is
also essential in determining the accretion history of the black hole in
question. The orbital parameters of the test cases are summarised in Table
1. Figure \ref{fig:1} shows the resulting light curves for the four tests
explored. The circularisation parameter is fixed at $\zeta = 0.1$ for all of
the tests.

We start by looking at the most luminous case, T1 in Table 1, shown with the
solid curve in Figure \ref{fig:1}. Low inclination and eccentricity of the
orbit ensure that the black hole spends all of its time inside the disc. In
addition, the relative velocity between the black hole and the gas is small as
the orbit of the former is close to circular. The capture rate is then large,
resulting in a relatively high gas capture rate and X-ray luminosity. The few
kinks in the lightcurve are caused by periodic variations in the relative
velocity due to the eccentricity of the orbit. One could analyse these in
terms of the epicyclic motion approximation for this nearly circular orbit.

The black hole in test T2 is on a slightly more eccentric and more inclined
orbit. The relative black hole-gas velocity is larger than in test T1, and
hence the gas capture rate (see eq. 1) is reduced.  The black hole spends a
significant amount of time inside the disc, but there are periods of time when
it exits the disc through its one or the other face.  Half of the dips in the
lightcurve correspond to time spent outside the disc, and the other to the
time when the relative velocity $\Delta v$ reaches the highest value along the
trajectory.  Due to a relatively large size of the accretion disc (the ``small
scale'' one discussed in \S \ref{sec:smalldisc}) that builds up around the
black hole, the viscous time is comparable to the duration of time spent
outside of the disc, and hence the dips are relatively minor. The X-ray light
curve of the source reaches a quasi-steady state with the luminosity $L_{\rm
x} \sim {\rm few} \times 10^{34}$ erg s$^{-1}$ after $\sim 10^4$ years.

In test T3 we consider a black hole on a more eccentric orbit, with
eccentricity $e = 0.6$, and a larger semi-major axis. The high eccentricity of
the orbit results in a high relative velocity which limits the gas capture
rate. For most of the orbit, $L_{\rm X} < 10^{33}$ erg s$^{-1}$. The
luminosity of the black hole increases dramatically at $t\approx 8000$ years,
when the black hole is at periapsis, where the relative velocity is low enough
and the density of the gas is high enough for the luminosity to approach that
of case T2.

Finally test T4 is close to the worst case scenario as far as the gas capture
rate is concerned. A high inclination and eccentricity orbit imply that the
black hole spends little time inside the disc. Accumulation of gas in the
small scale accretion disc happens in a burst-like manner when the black hole
is inside the disc. Also note that since the relative velocity is high, gas
capture radius (equation \ref{rcapt}) is smaller than it is in tests T1 and
T2, and hence the disc viscous time (equation \ref{tvisc}) is shorter,
resulting in shorter decay times for the ``bursts'' in the lightcurve.  The
X-ray luminosity is never larger than few~$\times 10^{32}$ erg s$^{-1}$.

These simple tests indicate that we may expect the black holes to produce a
detectable X-ray emission in one of the two ways: (i) a few black holes may be
on orbits essentially co-moving with the gas, producing a few bright point
sources; (ii) the dim majority of high inclination and/or high eccentricity
orbits may not produce individually bright sources but may be collectively
bright, producing an unresolvable ``diffuse'' X-ray emission.

\subsection{Representative cases}\label{sec:repr}

Having considered the individual accretors case in the previous section we
move on to the problem of the total black hole cusp emission with the
Minispiral model as described in \S \ref{sec:mini}.  The upper panels in
Figure 2 display the total X-ray luminosity of a cluster of 5000 black holes
as a function of time for two values of the circularisation radius parameter,
$\zeta$, $0.1$ and $0.001$, left and right, respectively. The lower panels
show the number of X-ray sources with luminosity higher than $10^{33}$ erg
s$^{-1}$ for the tests shown in the panels just above the respective lower
panels. Such sources could be observed by {\em Chandra}. Several conclusions
can be made. With a larger value of $\zeta=0.1$, the accretion discs around
black holes are larger, and thus viscous times are long. As a result, the
X-ray emission varies smoothly with time, first increasing as the discs are
built up, and then decreasing on $\sim$ a thousand years time scale. Thus the
sources are rather steady in time, and are also dim.

For the smaller value of $\zeta=10^{-3}$, viscous times in small scale discs
are much shorter. Therefore, the X-ray emission from the sources vary on much
shorter time scales, i.e., of few years to tens of years. The sources are also
brighter as the peak accretion rates are higher -- each individual source
shines much brighter for a shorter time, of course, as compared with the
larger $\zeta$ case. Both the upper and the lower panels provide us with
largely independent predictions which may be compared to X-ray observations.

Figure 3 shows the same experiments as Figure 2 but for 20000 black
holes. Comparison between the two different values of $\zeta$ shows similar
trends as before. It is interesting to compare the Figures 2 and 3. While the
results depend significantly on the a priori unknown value of $\zeta$, both
low and high $\zeta$ tests show same tendency of a significant luminosity
increase with increase in the number of black holes. In fact, the luminosity
increased by a larger factor than the black hole number did. The number of
sources above the chosen luminosity threshold also increased.  This suggests
that by performing tests across all reasonable parameter space for $\zeta$ we
should be able to find the maximum allowed number of stellar mass black holes
in the cusp.

\begin{figure}
\begin{center}
\epsfig{figure=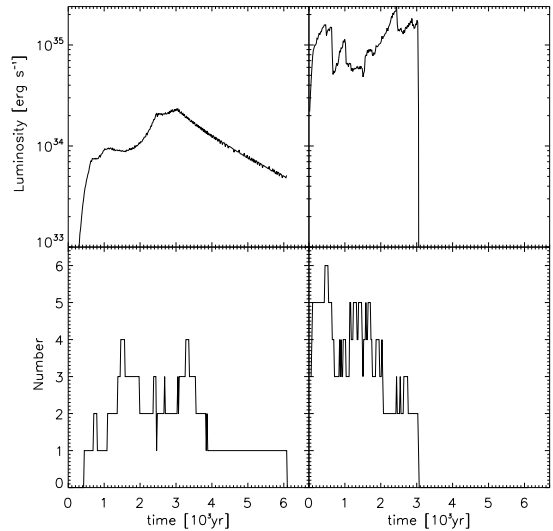}
\caption{X-ray light curves (top panels) and number of individual sources
  (bottom panels) where $L_{\rm x} < 10^{33}$ erg s$^{-1}$, when the total
  number of black holes in the inner parsec is 5000. The left and right panels
  correspond to $\zeta=0.1$ and $\zeta=0.001$ respectively.}
\label{fig:n5}
\end{center}
\end{figure}

\begin{figure}
\begin{center}
\epsfig{figure=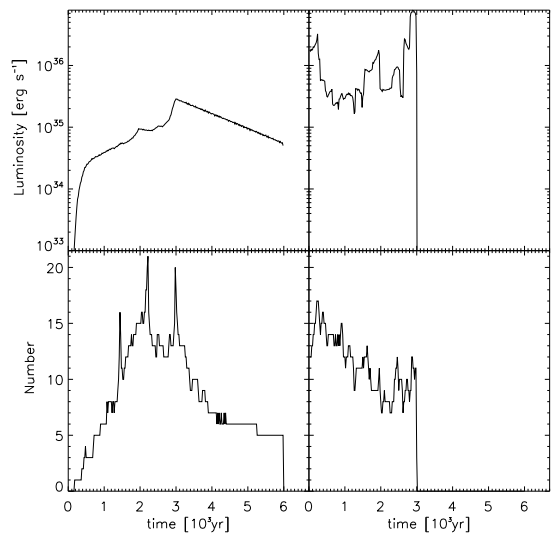}
\caption{X-ray light curves and number of individual sources where $L_{\rm x} <
  10^{33}$ erg s$^{-1}$, when the total number of black holes in the inner
  parsec is 20000. The left and right panels correspond to $\zeta=0.1$ and
  $\zeta=0.001$ respectively.}
\label{fig:n20}
\end{center}
\end{figure}

\subsection{Search in the \textbf{$N$}, \textbf{$\zeta$} parameter space}
\label{sec:search}

Following this idea, we ran a number of models for a range of values of
$\zeta$ and for the total black hole numbers of $N=5, 10, 20$~and~40 thousand.
During the time period modelled, the results vary considerably in each
test. For example, it is possible for just one single source to dominate the
X-ray luminosity output of the cluster.  In order to reduce and estimate
statistical noise of the results, for each of the values of $\zeta$ and the
total black hole number considered here, the tests were repeated three times,
each time generating a new random black hole orbit distribution. We then
calculate the mean value for the observables for the three runs, and we also
find deviations from the mean values. The averaging is done between time 
$2000 < t< 3000$ years as to look at a state that may be similar to the
present 
state of the Minispiral, given its estimated dynamical time.

A summary of the results is presented in Table 2. We chose to define several
quantities. The most important are the average total X-ray luminosity of the
black hole cluster and the number of black holes brighter than $10^{33}$ erg
s$^{-1}$, $N_{\rm X}$, as such sources would have been resolved by {\em
  Chandra} into 
separate point sources. Time-dependent variants of these quantities were
plotted in Figures 2 and 3.

In addition to these, we defined the probabilities that the total luminosity
of the cusp exceeds 10$^{35}$ and 10$^{36}$ erg s$^{-1}$ as the fraction of
time these conditions are satisfied:
\begin{equation}
P(L_X>10^{35})=\frac{1}{{\rm t_f}- {\rm t_i}}\;\int_{L_{\rm X} >
    10^{35}} \rm dt\;,
\label{p35}
\end{equation} 
and obviously similarly for 10$^{36}$ erg s$^{-1}$, $\rm t_i$ and $\rm t_f$
are 2000 and 3000 years respectively. 

The probability of the number of individual sources brighter than
 $\rm L_X>10^{33}$ being larger than 3, 10 and 20 at a given time is defined
 in a similar way, e.g., 
\begin{equation} P(N_X>10)=\frac{1}{{\rm t_f}- {\rm t_i}}\int_{
 N_{\rm X}>10}\rm 
    dt .
\end{equation} 
These values can be compared to the number of discreet X-ray sources in the
inner parsec as detected by {\em Chandra}.
    
Observations of the inner parsec by {\em Chandra} have placed upper limits on
the total luminosity of sources of approximately $10^{35}$ erg s$^{-1}$
\citep{Baganoff03a} and the number of individual X-ray sources with a
luminosity greater than $10^{33}$ erg s$^{-1}$ of a dozen or so (F. Baganoff,
private communication). With these constraints in mind we can immediately rule
out the possibility that the cusp contains 40000 black holes. For any
reasonable value of $\zeta$, the total luminosity and the number of individual
sources with $L_X > 10^{33}$ erg s$^{-1}$ are too large compared to
observations. A cusp containing 20000 black holes is not very likely but
cannot be ruled out completely at this time. In particular, only the larger
$\zeta$ case is acceptable for $N=$ 20000. Even though the average total cusp
luminosity is $\sim 10^{36}$ erg s$^{-1}$ for this test, i.e., too large, the
probability $P(L_X>10^{35})$ is only $\sim 0.66$.  Cusps with $N=$ 5000 or
10000 black holes is well within the limits imposed by observations.

\subsection{Sensitivity of results to the properties of the
  Minispiral}\label{sec:sens} 
           
We used a rather simple model for the Minispiral here (\S \ref{sec:mini}),
partially because it is not yet clear what a better model for this gas would
be. At the moment, we do not know the origin of this gaseous feature and the
precise three-dimensional distribution of gas and velocity field
\citep{Paumard04}.

To test sensitivity of our conclusions to the properties of the Minispiral, we
varied several of the assumptions made in \S \ref{sec:mini}. Table 3
summarises these tests. In particular, in one of the series of tests we let
the Minispiral to be 3 times more massive, i.e., to contain $150 \msun$ of
gas, with all other assumptions unchanged. In this case the luminosity of the
cusp increases significantly at a given number of black holes, and even the
$N=10^4$ case is too luminous for the smaller values of the circularisation
parameter $\zeta$. Hence the upper limit on the number of black holes is then
around $10^4$.

Another likely complication is that the gas may be on a parabolic or an
eccentric trajectory rather than a circular one, as assumed in this paper
until now. For such orbits, gas velocity can be both larger and smaller than
the local Keplerian value, depending on where exactly on the orbit the gas is.
Observationally, the Minispiral seems to be closer to the pericenter of its
orbit rather than an apocenter \citep{Paumard04}. To test the significance of
non-Keplerian orbits, we let the gas velocity to be 1.2 and 1.4 times the
local Keplerian value in the two series of tests presented in Table 3.
Clearly this model is not geometrically self-consistent as the half-disc we
use for the Minispiral should then deform in a complicated way, but we expect
to gain some guidance to the direction of change in the results nonetheless.
The larger gas velocity should result in a decrease in the number of black
holes travelling at a low relative speed through the Minispiral, which should
reduce the average gas capture rate (equation 1).

Table 3 shows that increasing the gas velocity to 1.2 of the local Keplerian
value results in a marked drop in average luminosity of black holes, to the
point that a cusp containing 40000 black holes cannot be ruled out for the
larger value of the circularisation parameter, $\zeta=0.1$. On the other hand,
realistically, we would expect the sources to have a distribution in values of
$\zeta$, and we would thus expect a fair number of sources to have $zeta \sim
0.01$ or less, which would then be ruled out.

Moving on to the gas velocity $= 1.4 v_{\rm K}$, we note a further drop in the
luminosity of the cusp and the number of sources detectable by {\em Chandra}
(Table 3). Even the $N=$ 40000 is allowed. We however feel that this model
strongly over-simplifies the situation in the Galactic Centre. The Minispiral
should be a feature bound to \sgra\ (or else the fact that it is crossing the
innermost region of the parsec now would be a coincidence), and hence it is
rather implausible that the gas is on a radial trajectory for which one would
have $v = \sqrt{2} v_{\rm K} \sim 1.4 v_{\rm K}$. We believe the case of $v=
1.2 v_{\rm K}$ is the one that represents the reality better.

Finally, the last entries in Table 3 are the tests with the Minispiral mass of
$M=150\msun$ and $v= 1.2 v_{\rm K}$. A cusp of 40000 black holes is clearly
inconsistent with the observations, whereas the $N=20000$ is constrained but
not completely ruled out.
       
\subsection{A neutron star cusp}
\label{sec:neutron}

The mechanism that produces an overabundance of black holes in the Galactic
Centre also applies to neutron stars, as they are also heavier than an average
star. Simulations by \cite{Freitag06} predict that a cusp of neutron stars
will have a number density profile quite similar to that of stellar mass black
holes. As far as our investigation goes, these neutron stars would be
accreting gas in a similar manner to the black holes, with modifications only
due to the smaller mass (we set $M_{\rm ns} =1.4 \msun$) and the existence of
a surface.

We approximate the emission of an accreting neutron star as a black body one
with temperature $T_{\rm ns}$ from the surface area $4\pi R_{\rm ns}^2$, where
$R_{\rm ns} =10$ km is the radius of the neutron star. Most of the radiation
flux will be emitted at wavelengths corresponding to photon energy $E= 3
kT_{\rm ns}$:
\begin{equation}
E \approx 0.8 ~~ L_{34}^{1/4}~~ KeV\;,
\end{equation} 
where $L_{34}$ is the X-ray Luminosity in units of $10^{34}$ erg s$^{-1}$. Due
to the quite large absorbing column density to the GC, $N_{\rm H} \sim
10^{23}$ cm$^{-2}$ \citep{Baganoff03a}, soft X-ray emission below $\sim 1$ keV
is practically unobservable. We hence set the minimum observable total X-ray
luminosity of a neutron star to be $10^{34}$ erg s$^{-1}$, rather than
$10^{33}$ erg s$^{-1}$ for the black hole case.

We assume that the radiative efficiency of an accreting neutron star with
negligible magnetic fields is constant at $\epsilon \sim 0.1$, in which case
the luminosity of an accreting neutron star is simply
\begin{equation}L_X = 0.1\dot{M}_{\rm acc}c^2\;.
\end{equation}
Note that if magnetic fields are non-negligible, then the ``propeller effect''
may reduce the X-ray luminosity of neutron stars \citep{MenouEtal99}, but we
leave these complications for a future more extended study.

With these modifications, we can use the machinery developed in \S
\ref{sec:method} to calculate the X-ray emission from the cusp containing
neutron stars. The results are presented in a way identical to the black hole
cusp in Table 4. As with the black hole cusp case, the $N=40000$ case is
strongly ruled out on account of too large a number of detectable point
sources, $N_{\rm X}$, and the total X-ray luminosity of the cusp. The
$N=20000$ cusp also appears to be a bit too high in terms of both the total
luminosity and the number of sources. The data may still tolerable this,
especially if one takes into account the possibility of the propeller effect
\citep{MenouEtal99} neglected here.

\section{Conclusions}
Stellar mass black holes and neutron stars are predicted to clutter the
central parsec of our Galaxy
\citep{Morris93,Miralda00,Freitag06,Hopman06}. While these predictions seem to
be very robust, observational confirmation of the existence of a stellar
remnant cusp is only indirect at the moment \citep{Muno05}. \cite{NS07}
suggested that these sources, accreting {\em cold} gas episodically from the
Minispiral or other molecular or ionised gas features found in the central
parsec, may be bright both collectively and individually to be observable with
{\em Chandra}. Here we presented a more elaborated study, where a
time-dependent disc accretion onto the compact sources was considered. Also,
we took into account the fact that at low accretion rates, the radiative
efficiency of black holes appears to be drastically reduced
\citep[e.g.,][]{Esin97}, and we used a Monte-Carlo like approach to randomly
initialise the cusp of compact remnants.

The main effort in our paper was to set the upper limit on the number of
compact remnants. Whereas the models have internal uncertainties, such as the
value of circularisation parameter $\zeta$, and observational uncertainties
(the Mass and precise orbit of the Minispiral), a cusp of black holes or
neutron stars with $N \simgt 40,000$ seems to be strongly ruled out. A cusp
with $N\sim 20,000$ black holes, as theoretically predicted
\citep{Freitag06,Hopman06}, is broadly consistent with the data. Future
efforts should improve these upper limits.

On the basis of our calculations, we think it is quite realistic that some of
the X-ray sources visible in the central parsec
\citep[e.g.,][]{Baganoff03a,Muno05} may be isolated black holes and neutron
stars accreting gas from the Minispiral. Such sources should be preferentially
found close to the Minispiral if viscous time is short ($\zeta$ is
small). In addition, binary systems containing a black hole and a normal low
mass star can also accrete gas in roughly the same way as we calculated
here. In the case of low values of circularisation parameter, $\zeta$, the
size of the disc around the primary (the black hole) can be smaller than the
size of the binary itself. Thus, these systems may appear as ``fake X-ray
binaries'', where the gas supply comes from outside rather than from the low
mass secondary.  Observational signatures of such systems might be warped and
out of binary plane accretion discs, ``too short'' or ``too weak'' accretion
outbursts for the size of the binary.

We acknowledge stimulating discussions with Mark Freitag, Fred Baganoff and
Rashid Sunyaev.

\begin{table*}
\caption{Characteristics of black hole cusp averaged between 2000--3000 years (see
  \S \ref{sec:search}).} 
\begin{center}
\begin{tabular}{|c|c|c|c|c|c|c|c|c|}\hline
Number$^a$ & $\zeta$$^b$ & $<L_{\rm X}>$$^c$ & P($L_{\rm X}>10^{36}$)$^d$ &
P($L_{\rm X}>10^{35}$)$^e$&$<N_X>$$^f$&P($N_{\rm X}>$20)$^g$&P($N_{\rm
  X}>$10)$^h$&P($N_{\rm X}>$3)$^i$\\

[$10^3$]&&[$10^{35}$erg s$^{-1}$]&&\\
                                  \hline
                                  \hline
5&	0.001&	1.85 $\pm$ 0.51&	0 &	0.53 $\pm$ 0.5&	2.76 $\pm$
                                  0.75&	0&	0.00 
$\pm$ 0.00&	0.33 $\pm$ 0.8\\ \hline
5&	0.01&	2.75 $\pm$ 1.33&    0.42 $\pm$ 0.2&	0.72 $\pm$ 0.1&	
4.2 $\pm$ 0.4&	0&	0&	0.63 $\pm$ 0.13\\ \hline
5&	0.1&	0.73 $\pm$ 0.6&	0&	0.24 $\pm$ 0.3&	2.23 $\pm$ 0.2&
0&	0&	0.18 $\pm$ 0.11\\ \hline
\hline
									
10&	0.001&	15.91 $\pm$ 2.70&	0.33 $\pm$ 0.13&	0.90 $\pm$ 0.07&	4.44 $\pm$ 0.28&	0&	0&	0.71 $\pm$ 0.06\\ \hline
10&	0.01&	1.03 $\pm$ 0.24&	0&	0.45 $\pm$ 0.16&	5.36
$\pm$ 0.26&	0&	0 &	0.90 $\pm$ 0.04\\ \hline
10&	0.1&	0.51 $\pm$ 0.28&	0&	0.31 $\pm$ 0.22&	4.53 $\pm$ 1.52&	0&	0.01 $\pm$ 0.00&	0.45 $\pm$ 0.21\\ \hline
\hline									
20&	0.001&	19.97 $\pm$ 5.87&	0.57 $\pm$ 0.05&	0.99 $\pm$
0.01&	8.87 $\pm$ 0.81&	0&	0.19 $\pm$ 0.11&	1.00 $\pm$ 0.00\\ \hline
20&	0.01&	9.67 $\pm$ 1.94&	0.34 $\pm$ 0.14&	0.99 $\pm$ 0.00&
12.26 $\pm$ 1.40&	0.01 $\pm$ 0.01&	0.69 $\pm$ 0.17&	1.00 $\pm$ 0.00\\ \hline
20&	0.1&	9.88 $\pm$ 4.49&	0.34 $\pm$ 0.22&	0.66 $\pm$ 0.23&	13.11 $\pm$ 1.25&	0.01 $\pm$ 0.00&	0.74 $\pm$ 0.16&	1.00 $\pm$ 0.00\\ \hline
\hline									
40&	0.001&	40.37 $\pm$1.75&	1.00 $\pm$ 0.00&	1.00 $\pm$ 0.00&	19.40 $\pm$ 0.37&	0.34 $\pm$ 0.05&	1.00 $\pm$ 0.00&	1.00 $\pm$ 0.00\\ \hline
40&	0.01&	100.53 $\pm$ 62.13&	0.77 $\pm$ 0.08&	0.99 $\pm$ 0.00&	24.66 $\pm$ 0.50&	0.85 $\pm$ 0.05&	1.00 $\pm$ 0.00&	1.00 $\pm$ 0.00\\ \hline
40&	0.1&	6.50 $\pm$1.83&	0.20 $\pm$ 0.14&	1.00 $\pm$ 0.00&	24.59 $\pm$ 2.18&	0.72 $\pm$ 0.14&	1.00 $\pm$ 0.00&	1.00 $\pm$ 0.00\\ \hline
\end{tabular}
\end{center}
\footnotesize{
\leftline{The collumns list:}

\leftline{$^a$ Total number of black holes in the cusp}

\leftline{$^b$ Circularisation parameter (\S \ref{sec:smalldisc})}

\leftline{$^c$  Time-averaged luminosity of the cusp}

\leftline{$^{d-e}$ Probability that the total luminosity of the cusp is greater
    than $10^{36}$ or $10^{35}$ erg s$^{-1}$, respectively.}


\leftline{$^f$ Average number of sources with X-ray luminosity
    greater than $10^{33} $erg s$^{-1} (N_{\rm X})$ }

\leftline{$^{g-i}$ Probability that $N_{\rm X}$ is greater than 20, 10 and 3, respectively} 

}
\end{table*}

\begin{table*}
\caption{Same as Table 2, but for different models of the Minispiral}
\begin{center}
\begin{tabular}{|c|c|c|c|c|c|c|c|c|}\hline
Number & $\zeta$ & $<L_X>$ & P($L_{\rm X}>10^{36}$) &
P($L_X>10^{35}$)&$<N_{\rm X}>$&P($N_{\rm X}>$20)&P($N_{\rm X}>$10)&P($N_{\rm
  X}>$3)\\ 

[$10^3$]&&[$10^{35}$erg s$^{-1}$]&&\\
                                  \hline
                                  \hline
3 Mgas$^a$	\\							
5&	0.01&	28.13 $\pm$ 11.54&	0.62 $\pm$ 0.15&	0.84 $\pm$ 0.08&	5.81 $\pm$ 0.95&	0&	0.03 $\pm$ 0.02&	0.84 $\pm$ 0.06\\\hline
5&	0.1&	1.53 $\pm$ 4.45&	0&	0.67 $\pm$ 0.40&	7.04 $\pm$ 0.47&	0&	0.06 $\pm$ 0.04&	1.00 $\pm$ 0.00\\\hline
10&	0.01&	35.21 $\pm$ 12.01&	0.63 $\pm$ 0.17&	1.00 $\pm$ 0.00&	9.24 $\pm$ 0.93&	0&	0.36 $\pm$ 0.13&	1.00 $\pm$ 0.00\\\hline
10&	0.1&	2.06 $\pm$ 0.74&	0&	0.56 $\pm$ 0.16&	14.12 $\pm$ 0.51&	0.02 $\pm$ 0.00&	0.98 $\pm$ 0.01&	1.00 $\pm$ 0.00\\\hline
20& 0.01& 24.53 $\pm$ 6.02& 0.77 $\pm$ 0.13 & 1.00 $\pm$ 0.00& 18.95 $\pm$
                                  1.25& 0.40 $\pm$ 0.16&1.00 $\pm$ 0.00& 1.00 $\pm$ 0.00\\\hline
20&0.1& 5.47 $\pm$ 0.41& 0.01 $\pm$ 0.00& 1.00 $\pm$ 0.00& 29.38 $\pm$ 2.06&
                                  0.90 $\pm$ 0.07& 1.00 $\pm$ 0.00& 1.00 $\pm$ 0.00\\\hline
\hline								

$1.2 v_{\rm K}^b$\\								
5&	0.01&	0.93 $\pm$ 0.55&	0.02 $\pm$ 0.01&	0.14 $\pm$ 0.07&	1.85 $\pm$ 0.36&	0&	0&	0.10 $\pm$ 0.04\\\hline
5&	0.1&	0.05 $\pm$ 0.03&	0&	0&	0.44 $\pm$ 0.20&	0&	0&	0\\\hline
10&	0.01&	0.45 $\pm$ 0.10&	0&	0.13 $\pm$ 0.05&	3.47 $\pm$ 0.42&	0&	0&	0.49 $\pm$ 0.13\\\hline
10&	0.1&	0.38 $\pm$ 0.25&	0&	0.17 $\pm$ 0.12&	2.45 $\pm$ 1.42&	0&	0&	0.31 $\pm$ 0.22\\\hline
20&	0.01&	4.86 $\pm$ 1.63&	0.14 $\pm$ 0.05&	0.43 $\pm$ 0.15&	6.65 $\pm$ 0.08&	0&	0.03 $\pm$ 0.01&	0.91 $\pm$ 0.04\\\hline
20&	0.1&	0.35 $\pm$ 0.07& 	0&	0&	4.24 $\pm$ 0.08&	0&	0&	0.76 $\pm$ 0.07\\\hline
40&	0.01&	40.65 $\pm$ 1.66&	0.56 $\pm$ 0.14&	0.94 $\pm$ 0.04&	13.42 $\pm$ 0.23&	0.02 $\pm$ 0.01&	0.74 $\pm$ 0.06&	1.00 $\pm$ 0.00\\\hline
40&	0.1&	1.05 $\pm$ 0.37 &	0&	0.18 $\pm$ 0.13&	10.27 $\pm$ 0.62&	0&	0.44 $\pm$ 0.80&	1.00 $\pm$ 0.00\\\hline
\hline
1.4 $v_{\rm K}^c$	\\							
5&	0.01&	0.07 $\pm$ 0.02&	0&	0&	0.71 $\pm$ 0.15&	0&	0&	0\\\hline
5&	0.1&	0.02 $\pm$ 0.00&	0&	0&	0.37 $\pm$ 0.26&	0&	0&	0\\\hline
10&	0.01&	0.12 $\pm$ 0.04&	0&	0&	1.17 $\pm$ 0.21&	0&	0&	0\\\hline
10&	0.1&	0.05 $\pm$ 0.01&	0&	0&	0.83 $\pm$ 0.32&	0&	0&	0\\\hline
20&	0.01&	0.32 $\pm$ 0.48&	0&	0.03 $\pm$ 0.02&	3.66 $\pm$ 0.43&	0&	0&	0.49 $\pm$ 0.13\\\hline
20&	0.1&	0.03 $\pm$ 0.00&	0&	0&	0.11 $\pm$ 0.05&	0&	0&	0\\\hline
40&	0.01&	0.64 $\pm$ 0.14&	0&	0.18 $\pm$ 0.09&	5.37 $\pm$ 0.60&	0&	0.04 $\pm$ 0.01&	0.72 $\pm$ 0.07\\\hline
40&	0.1&	0.12 $\pm$ 0.03&	0&	0&	1.34 $\pm$ 0.36&	0&	0&	0.01 $\pm$ 0.01\\\hline
\hline								
3 Mgas 	\&&   1.2 $v_{\rm K}^d$\\	
20&	0.01&	9.79 $\pm$ 0.93&	0.34 $\pm$ 0.06&	1.00 $\pm$ 0.00&	15.47 $\pm$ 0.25&	0.07 $\pm$ 0.02&	0.98 $\pm$ 0.01&	1.00 $\pm$ 0.00\\\hline
20&	0.1&	2.89 $\pm$ 1.03&	0.09 $\pm$ 0.06&	0.86 $\pm$ 0.10&	13.72 $\pm$ 0.25&	0.04 $\pm$ 0.01&	0.91 $\pm$ 0.03&	1.00 $\pm$ 0.00\\\hline
40&	0.01&	12.17 $\pm$ 0.96&	0.59 $\pm$ 0.02&	1.00 $\pm$ 0.00&	25.02 $\pm$ 0.78&	0.90 $\pm$ 0.05&	1.00 $\pm$ 0.00&	1.00 $\pm$ 0.00\\\hline
40&	0.1&	6.43 $\pm$ 1.76&	0.29 $\pm$ 0.10&	0.93 $\pm$ 0.05&	19.63 $\pm$ 1.82&	0.50 $\pm$ 0.16&	1.00 $\pm$ 0.00&	1.00 $\pm$ 0.00\\\hline

\end{tabular}
\end{center}
\footnotesize{
\leftline{$^a$ Mass of the Minispiral has been tripled to 150$\msun$ }

\leftline{$^b$ Velocity of the gas has been increased to 1.2~$v_{\rm K}$, where
  $v_{\rm K}$ is
  the local keplerian velocity}

\leftline{$^c$ Velocity of the gas has been increased to 1.4~$v_{\rm K}$ } 

\leftline{$^d$ Mass of the Minispiral has been tripled to 150$\msun$ and the
  velocity of the gas has been increased to 1.2~$v_{\rm K}$ }

}

\end{table*}

\begin{table*}
\caption{Same as Table 2 but for a neutron star cusp.}
\begin{center}
\begin{tabular}{|c|c|c|c|c|c|c|c|c|}\hline
Number$^a$ & $\zeta$$^b$ & $<L_{\rm X}>$$^c$ & P($L_{\rm X}>10^{36}$)$^d$ &
P($L_X>10^{35}$)$^e$&$<N_{\rm X}>$$^f$&P($N_{\rm X}>$20)$^g$&P($N_{\rm
  X}>$10)$^h$&P($N_{\rm X}>$3)$^i$\\ 

[$10^3$]&&[$10^{35}$erg s$^{-1}$]&&\\
                                  \hline
                                  \hline
5&	0.001&	3.23 $\pm$ 5.24&	0.08 $\pm$ 0.06&	0.79 $\pm$ 0.11&	2.84 $\pm$ 0.43&	0&	0&	0.32 $\pm$ 0.10\\ \hline
5&	0.01&	2.43 $\pm$ 0.76&	0&	0.67 $\pm$ 0.12&	2.78 $\pm$ 0.81&	0&	0&	0.38 $\pm$ 0.16\\ \hline
5&	0.1&	1.44 $\pm$ 0.18&	0&	0.85 $\pm$ 0.11&	2.90 $\pm$ 0.67&	0&	0&	0.34 $\pm$ 0.20\\ \hline
\hline								
10&	0.001&	4.17 $\pm$ 0.90&	0.02 $\pm$ 0.01&	0.99 $\pm$ 0.01&	6.25 $\pm$ 0.45&	0&	0.03 $\pm$ 0.02&	0.94 $\pm$ 0.02\\ \hline
10&	0.01&	5.89 $\pm$ 1.38&	0.18 $\pm$ 0.08&	0.99 $\pm$ 0.01&	4.53 $\pm$ 0.34&	0&	0&	0.72 $\pm$ 0.10\\ \hline
10&	0.1&	6.69 $\pm$ 3.59&	0.30 $\pm$ 0.21&	0.83 $\pm$ 0.12&	3.32 $\pm$ 0.15&	0&	0&	0.47 $\pm$ 0.05\\ \hline
\hline								
20&	0.001&	8.13 $\pm$ 1.34&	0.27 $\pm$ 0.11&	1.00 $\pm$ 0.00&	11.52 $\pm$ 1.15&	0&	0.57 $\pm$ 0.15&	1.00 $\pm$ 0.00\\ \hline
20&	0.01&	9.73 $\pm$ 1.73&	0.32 $\pm$ 0.12&	1.00 $\pm$ 0.00&	7.32 $\pm$ 0.30&	0&	0.11 $\pm$ 0.08&	0.97 $\pm$ 0.02\\ \hline
20&	0.1&	7.17 $\pm$ 2.15&	0.27 $\pm$ 0.19&	1.00 $\pm$ 0.00&	9.00 $\pm$ 0.35&	0&	0.19 $\pm$ 0.05&	1.00 $\pm$ 0.00\\ \hline
\hline								
40&	0.001&	41.60 $\pm$ 12.71&	0.99 $\pm$ 0.01&	1.00 $\pm$
0.00&	21.99 $\pm$ 0.52&	0.69 $\pm$ 0.07&	1.00 $\pm$ 0.00&
1.00 $\pm$ 0.00\\ \hline 
40&	0.01&	18.62 $\pm$ 4.46&	0.87 $\pm$ 0.07&	1.00 $\pm$
0.00&	15.82 $\pm$ 0.48&	0.04 $\pm$ 0.02&	0.99 $\pm$ 0.00&
1.00 $\pm$ 0.00\\ \hline 
40&	0.1&	17.91 $\pm$ 3.74&	0.86 $\pm$ 0.10&	1.00 $\pm$
0.00&	21.34 $\pm$ 0.34& 0.66 $\pm$ 0.05&	1.00 $\pm$ 0.00&
1.00 $\pm$ 0.00\\ \hline 

\end{tabular}
\end{center}
\footnotesize{
\leftline{$^a$ $^-$ $^e$ see Table 2 caption.}

\leftline{$^f$ Average number of neutron stars with L$_{\rm X}$
    greater than $10^{34} $erg s$^{-1} (N_{\rm X})$. Only sources with L$_{\rm X}$ $> 10^{34}$ are
    visible, see \S \ref{sec:neutron}. }

\leftline{$^g$ $^-$ $^i$ see Table 2 caption.}}
\end{table*}

\bibliographystyle{mnras} 

\end{document}